\documentclass[twocolumn,showpacs,preprintnumbers,amsmath,amssymb,amsfonts]{revtex4-1}

\usepackage[T1]{fontenc}
\usepackage{color}
\usepackage{graphicx}
\usepackage{palatino}
\usepackage{mathpazo}
\usepackage{amsmath}
\usepackage{bm}

\begin{document}

\title{Effect of pressure on the number of dynamically correlated molecules when approaching the glass transition}

%\classification{<Replace this text with PACS numbers; choose from this list:
%                \texttt{http://www.aip..org/pacs/index.html}>}
%\keywords      {<Enter Keywords here>}

\author{Christiane Alba-Simionesco}
\affiliation{Laboratoire L\'eon Brillouin, UMR 12, CEA-CNRS, 91191 Saclay, France}

\author{C{\'e}cile Dalle-Ferrier}
\affiliation{Laboratoire L\'eon Brillouin, UMR 12, CEA-CNRS, 91191 Saclay, France}

\author{Gilles Tarjus}
\affiliation{LPTMC, CNRS-UMR 7600, Universit\'e Pierre et Marie Curie, 4 Pl. Jussieu, 75252 Paris c\'edex 05, France}

\begin{abstract}

We characterize the heterogeneous character of the dynamics of liquids approaching
the glass transition through an experimental determination of the number of dynamically
correlated molecules $N_{corr}$ as obtained from dynamical susceptibilities. To do so, we have
obtained a new set of dielectric spectroscopy data for liquid dibutyl-phtalate on
a fine and extended temperature and pressure grid, and we have used it in conjunction with high-pressure
data from the literature.  We have been able to evaluate the contributions to $N_{corr}$  that are due to
fluctuations associated with density and with temperature separately,
thereby improving the estimate of $N_{corr}$.
We find that $N_{corr}$  increases along the glass transition line, and more generally along any
isochronic line, as pressure increases (up to $1\,GPa$), a result which is at
odds with recent reports and theoretical predictions.

\end{abstract}

\maketitle

\section{Introduction}

A tempting explanation of the dramatic slowing down of the
dynamics observed in liquids approaching their glass transition is
the existence of a growing underlying length scale. The relevance of such a length,
characterizing the cooperative nature of the structural relaxation, was already
suggested by Adam and Gibbs in the sixties \cite{Adam65}.  The search for length
scales went in two directions \cite{Tarjus11}: on the one hand, studies of nontrivial spatial correlations
in the structure, associated with some hidden order parameter or in the form of
point-to-set correlations describable through the influence of amorphous
boundary conditions \cite{PTS} (both being essentially inaccessible in laboratory experiments); on
the other hand, based on the mounting evidence in experiments and simulations of
the increasingly heterogeneous character of the dynamics \cite{DHexpts,Hurley95,DHbook},
investigations of a ``dynamic'' length describing
the spatial extent of the dynamic heterogeneities. However, direct experimental measurement
of  the latter in molecular liquids is hard to obtain. In consequence, the
corresponding results are scarce  \cite{DHexpts,Tracht98, Reinsberg},
moreover providing no information on the evolution with temperature.

A few years ago, on the basis of theoretical arguments, a new method for estimating the
typical length scale of the dynamical heterogeneities and access its
temperature dependence was suggested \cite{Berthier05}. It relies on the
introduction and the study of multi-point dynamical susceptibilities. No information on spatial correlations
in the dynamics can indeed be derived from the mere consideration of standard 2-time
correlation functions, $F(t)=<\delta O(0)\delta O(t)>$
(with $O(t)$ some observable depending on the probe), which describe the ``average'' dynamics of a
liquid. Higher-order space-time correlation functions, or their volume integrals that represent susceptibilities
and may therefore be more easily accessible, are required to probe the fluctuations around
the average dynamics,  whose manifestation is precisely the dynamic heterogeneities \cite{Franz99}. From the
multi-point dynamic susceptibilities, one can extract an estimate of the ``number of dynamically
correlated molecules'' \cite{Berthier05,Dalle07}, which is the focus of the present work.

Specifically, the four-point dynamical susceptibility $\chi_4(t)=N<\delta C(0,t)^2>$, where $N$ is
the number of molecules in the system and $\delta C(0,t) = \delta O(0)\delta O(t) - F(t)$ characterizes
deviations from the average dynamics, quantifies the overall amount of spatial correlations between
\emph{spontaneous} fluctuations taking place between times 0 and $t$. This quantity is still, unfortunately,
very difficult to directly measure in experiments, especially in the case of molecular liquids. Interestingly, however,
it is related through fluctuation-dissipation relations \cite{Berthier05,Berthier07} to other
dynamical susceptibilities associated with fluctuations that are \emph{induced} by an external control
parameter $x$ (such as temperature, pressure or,
for colloids, volume fraction):  these are three-point susceptibilities, defined as
$\chi_x(t)=\frac{\partial F(t)}{\partial x}$, which describe the
response of the two-time correlation function $F(t)$ to a change in an external parameter.
Such quantities are much easier to experimentally access than $\chi_4(t)$, as the
dependence of $F(t)$ on temperature, pressure, or any external field,  may be measured by
several well-developed methods such as dielectric or photon correlation spectroscopy for example \cite{Berthier05}.
It was moreover shown by computer simulation studies \cite{Berthier07} that a good
estimate for  $\chi_4(t)$ is obtained from the $\chi_x(t)$'s in the viscous liquid regime. (Note
that another 3-point dynamical susceptibility has also been directly measured by means of an investigation
of the nonlinear dielectric response \cite{Ladieu}.)

Thanks to these high-order dynamical susceptibilities, direct
experimental evidence of the increase of the number of dynamically correlated
molecules $N_{corr}(T)$ while approaching the glass transition by cooling was obtained for
several molecular liquids and polymers \cite{Dalle07,Dalle08,Capaccioli08,Fragiadakis09,Fragiadakis11,Ladieu}.
The growth of $N_{corr}(T)$ was found
to be moderate compared to the dramatic change in relaxation time in the same
temperature range, with $N_{corr}$ reaching a value of the order of $10^2$ at the glass transition
temperature $T_g$ at atmospheric pressure.
No clear conclusions could be drawn concerning the influence of the
chemical nature of the system on the variation of $N_{corr}$. In particular, no correlation
was observed between the growth of $N_{corr}(T)$ and the fragility of the
system, \emph{i.e.} the deviation of the temperature dependence
of relaxation time from an Arrhenius behavior.

When cooling a liquid at constant pressure, its density increases. It is therefore useful to
disentangle the effect coming from the densification of the system from that
due to the decrease of thermal energy. This provides clearer insight into the physics of the slowing
down of relaxation \cite{Ferrer98}. In addition, when considering estimates of
the four-point dynamical susceptibility $\chi_4(t)$, it may prove important because a better
description of the associated fluctuations in the $NPT$ ensemble is provided by summing the separate
contributions of the density-induced and the temperature-induced fluctuations. However, except
for partial results at atmospheric pressure \cite{Dalle07}, this latter procedure was never attempted. Previous
pressure studies along several isobars considered the crudest bound on $\chi_4(t)$ only \cite{Fragiadakis09,Fragiadakis11}.

In this work, we consider the effect of density and temperature on the number of dynamically correlated
molecules in a fragile glass-forming liquid, dibutylphtalate. To do so, we have measured the (linear)
dielectric susceptibility along both isobars and isotherms, covering thermodynamic  points on a fine grid in
temperature and pressure, and combined these results with existing high-pressure
data \cite{Cook93,Paluch03}. We have calculated the responses of the linear dielectric susceptibility
to temperature along isobars  and to density along isotherms to provide estimates of the variation
of the number of dynamically correlated molecules. To go beyond this and, as explained above, improve the estimate
of $N_{corr}$ by considering separately the contributions associated
with density and temperature at any state point, we have
used the density-temperature scaling already shown to describe relaxation data in supercooled liquids in a large domain of
temperature and density \cite{Alba02,Alba04,Roland04,Dreyfus04,Roland05}. With the help of this scaling
description, one can more easily compute the various dynamic quantities
entering in the expression of $N_{corr}$. We have taken special care to minimize
and to evaluate the systematic errors that come with the calculation of $N_{corr}$. Our main finding is that $N_{corr}$
significantly increases (by a factor of almost 3) along the glass transition line as one increases the pressure up to $1\,GPa$
(the density changes by $25\,\%$). The
same trend is found for other isochronic lines (\textit{i.e.} lines of equal relaxation time). Contrary to previous
reports  \cite{Fragiadakis09,Fragiadakis11} and to theoretical predictions \cite{Xia00,Lubchenko07},
the extent of the spatial correlations in the dynamics is therefore not uniquely determined by the relaxation time.

\section{Experimental details}

\subsection{Sample and experimental setup}

\begin{figure}%
\includegraphics*[width=\linewidth]{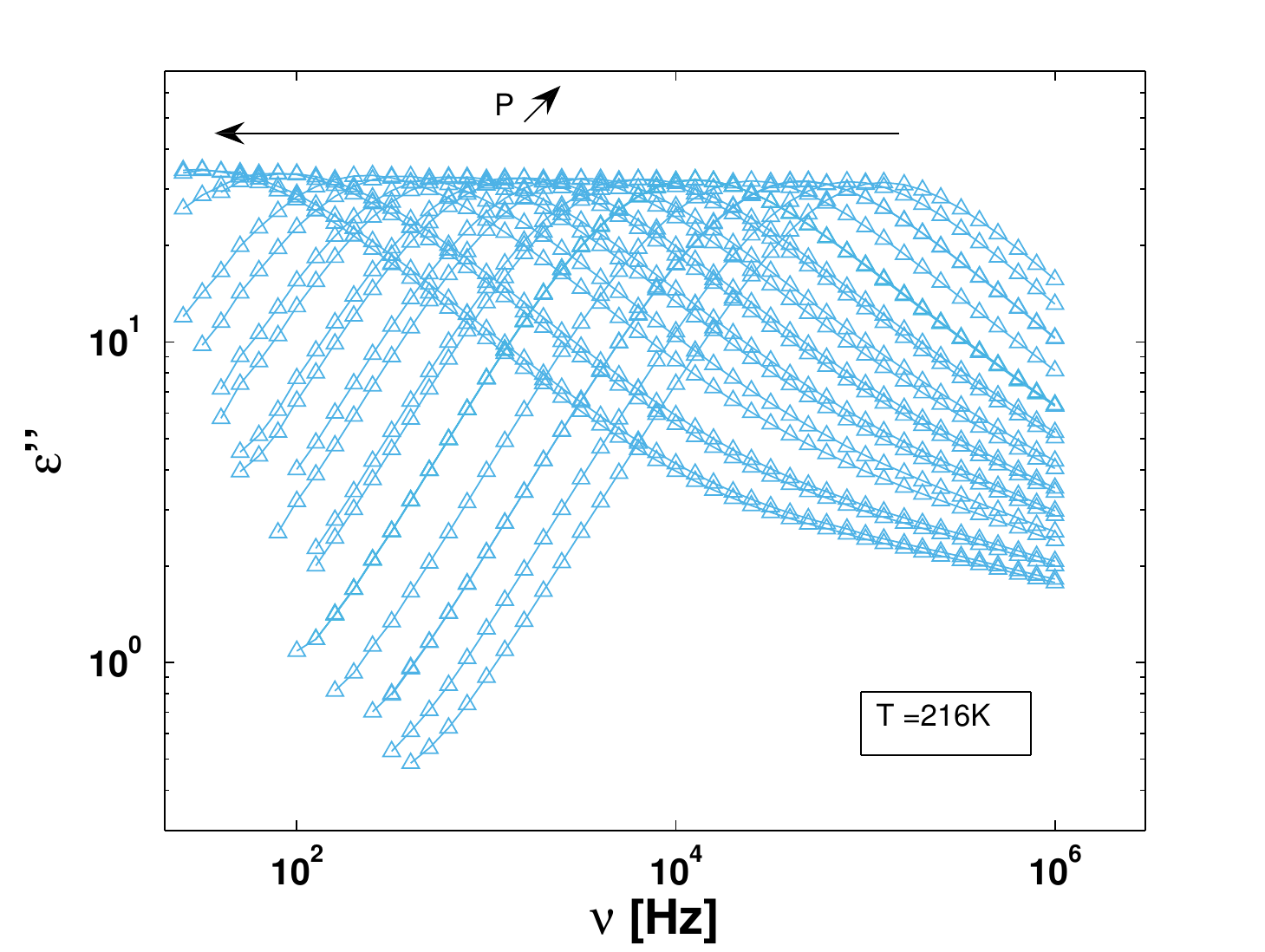}
\caption{\label{fig1}  Imaginary part of the dielectric spectrum of liquid DPB measured at $T=216\,K$
from atmospheric pressure (leftmost curve) to $P=250\,MPa$ (rightmost curve) obtained by both compression
and decompression. The average pressure step between 50 and 250 $MPa$ is about $10\,MPa$.}
\end{figure}

The dibutylphtalate (DBP) sample ($99\%$ purity) was acquired from
Sigma Aldrich and used as purchased. The experimental pressure dielectric setup for the dielectric
measurements was described in a previous work \cite{Niss07}.
The main characteristic of this setup is that it insures a completely
isotropic compression by the use of an external pressure fluid. For the present study, we
measured a new set of data with very small pressure or temperature steps (see Fig. \ref{fig1} for an illustration).
This data set is fully compatible with that reported in Ref. [\onlinecite{Niss07}] but is more complete
in order to perform the detailed analysis involved in estimating $N_{corr}$.  The reproducibility was always
checked very carefully, especially under pressure where the
experiments were carried out alternatively by compression and
decompression for comparison.

The dielectric spectra were fitted by Havriliak-Negami expressions \cite{Havriliak67}:
\begin{equation}
\label{eq:havriliak}
\epsilon(\omega)=\epsilon_{\infty} + \frac{\Delta \epsilon}{[1+(i\omega \tau)^{\alpha}]^{\gamma}}\;,
\end{equation}
with $\Delta \epsilon$, $\tau$, $\alpha$ and $\gamma$ temperature and pressure dependent adjustable
parameters. This provided a convenient
and faithful parametrization of all data, allowing for instance to conveniently extract the relaxation function
through Fourier transformation and to take numerical derivatives with respect to the control parameters via
the dependence of the adjustable parameters.

\subsection{Thermodynamic data}

Information on the thermodynamics is needed for the computation of the
number of dynamically correlated molecules and it must be considered with care in order to
minimize the sources of uncertainty. The PVT data was obtained from the literature
and included very high pressure results \cite{Cook94,Cook93,Bridgman32}. (This has been
already described in Ref. [\onlinecite{Niss07}].) The density and temperature dependent isothermal
compressibility $\kappa_T(T,\rho)$ and the thermal expansion coefficient $\alpha_P(T,\rho)$
were estimated from this data set.

Some results for the temperature dependence of the heat capacity $C_P$ of DBP are available in
the literature \cite{Mizukami95,Krueger99}, but only at
atmospheric pressure. We estimated from thermodynamic arguments and
comparisons to other molecular liquids
that the pressure variation of $C_P$ is negligible for DBP. This is in part due to the fact
that  the absolute value of $C_P$ is very high for DBP (about $440\,JK^{-1}mol^{-1}$ at $T_g$
and atmospheric pressure \cite{Mizukami95}).
The variation for the constant volume heat capacity $C_V$ can in turn be estimated from the relation
$C_V = C_P - T\alpha_P^2\,V/\kappa_T$ and is also found to be small (typically less than $10\,\%$).
The jump of $C_P$ at $T_g$ is known at atmospheric
pressure \cite{Mizukami95} and is about $150\,JK^{-1}mol^{-1}$. Unfortunately, no data
are available for the jump of $C_V$ nor for that of $\kappa_T$ at the
glass transition. (No data either are available for the pressure dependence of these jumps.)
For other molecular glass-forming liquids, one observes that the heat capacity
$C_P$ of the liquid at $T_g$ slightly increases with pressure: a few $\%$ for 3-methylpentane and 1-propanol
up to $200\,MPa$ \cite{Takahara94}, a few $\%$ for m-fluoroaniline up to $400\,MPa$ \cite{Alba91}, and $10-12\,\%$
for toluene up to $400\,MPa$ \cite{Ter88}. However, the $C_P$ of the glass at $T_g$ has
been found to also increase in 3-methylpentane
and 1-propanol where it has been measured \cite{Takahara94}, so that the heat capacity jump $\Delta C_{P,g}( P)$
barely changes with pressure in these systems.

\section{Number of dynamically correlated molecules $N_{corr}$ along isobars and isotherms}

We are interested in evaluating the four-point dynamical susceptibility $\chi_4$ and the associated
number of dynamically correlated molecules in the $NPT$ ensemble. From considerations about
fluctuation-dissipation relations \cite{Berthier05,Berthier07},  two expressions may be used.
The first one,
\begin{equation}
\label{eq:eqNcorrP}
\chi_4^{NPT}(t)=\frac{k_BT^2}{c_P}[\chi_T^{NPT}(t)]^2+\chi_4^{NPH}(t)\, ,
\end{equation}
where $c_P$ is the isobaric heat capacity per molecule, relates $\chi_4^{NPT}(t)$ to
the three-point-dynamical susceptibility
$\chi_T^{NPT}(t)=\left(\frac{\partial F(t)}{\partial T}\right)_P$, with a strictly positive residual term describing
the fluctuations in the $NPH$ ensemble (where $H$ is the enthalpy).
The second expression takes into account  the contributions associated
with temperature (at constant volume) and density (at constant temperature) separately:
\begin{equation}
\begin{split}
\chi_4^{NPT}(t)=\frac{k_BT^2}{c_V}[\chi_T^{NVT}(t)]^2
+\rho^3 k_B T\kappa_T [\chi_\rho^{NPT}(t)]^2+\chi_4^{NVE}(t)\, ,
\end{split}
\label{eq:eqNcorrrhoT}
\end{equation}
where $c_V$ is the isochoric heat capacity per molecule, $\chi_T^{NVT}(t)=\left(\frac{\partial F(t)}{\partial T}\right)_\rho$,
$\chi_{\rho}^{NVT}(t)=\left(\frac{\partial F(t)}{\partial \rho}\right)_T$, and $\chi_4^{NVE}$ is
the (strictly positive) four-point-dynamical susceptibility in the $NVE$ ensemble. For a more detailed discussion
concerning these different quantities, see Refs. [\onlinecite{Berthier07,Dalle07}].

In simulation studies on model glass-formers \cite{Berthier07}, $\chi_4^{NPH}$ and
$\chi_4^{NVE}$ were found to be small compared to the other terms in the viscous liquid regime
at low temperature, becoming even negligible at the lowest temperatures. This was also indirectly
confirmed by a comparison between results obtained from
$(k_B T^2/c_P)[\chi_T^{NPT}(t)]^2$ and from the direct experimental measurement of a nonlinear
dielectric susceptibility \cite{Ladieu}. It should nonetheless be stressed that, since $\chi_4^{NVE} \leq \chi_4^{NPH}$,
considering the separate contributions associated with temperature and density,
$(k_BT^2/c_V)[\chi_T^{NVT}(t)]^2+\rho^3k_BT\kappa_T [\chi_\rho^{NPT}(t)]^2$,
provides a better approximation to the dynamical
four-point susceptibility than $(k_B T^2/c_P)[\chi_T^{NPT}(t)]^2$.

In this work we focus on the numbers of dynamically correlated molecules
$N_{corr}$ which may be defined as the maximum
over time of the relevant time-dependent susceptibilities (or their absolute value if negative).
This maximum takes place for a time of the order of the average relaxation time of
the system $\tau_\alpha$. One therefore considers $N_{corr,4}^{NPT}=max_t[\chi_4^{NPT}(t)]$,
$N_{corr,T}^{NPT}=\sqrt{k_B/c_P}\,T max_t\{\vert \chi_T^{NPT}(t)\vert \}$,
$N_{corr,T}^{NVT}=\sqrt{k_B/c_V}\,T max_t\{\vert \chi_T^{NVT}(t)\vert \}$,
and $N_{corr,\rho}^{NPT}=\sqrt{\rho^3 k_B T \kappa_T} max_t\{\chi_{\rho}^{NPT}(t)\}$.
From Eqs. (\ref{eq:eqNcorrP}) and (\ref{eq:eqNcorrrhoT}), $N_{corr,4}^{NPT}$ may then be estimated either from
\begin{equation}
\label{eq:eqNcorrP_N}
N_{corr,4}^{NPT} \gtrsim (N_{corr,T}^{NPT})^2
\end{equation}
or from
\begin{equation}
\label{eq:eqNcorrrhoT_N}
N_{corr,4}^{NPT} \gtrsim (N_{corr,T}^{NVT})^2+ (N_{corr,\rho}^{NPT})^2,
\end{equation}
with the latter expression providing a better approximation than the former one.

\begin{figure}%
\includegraphics*[width=\linewidth]{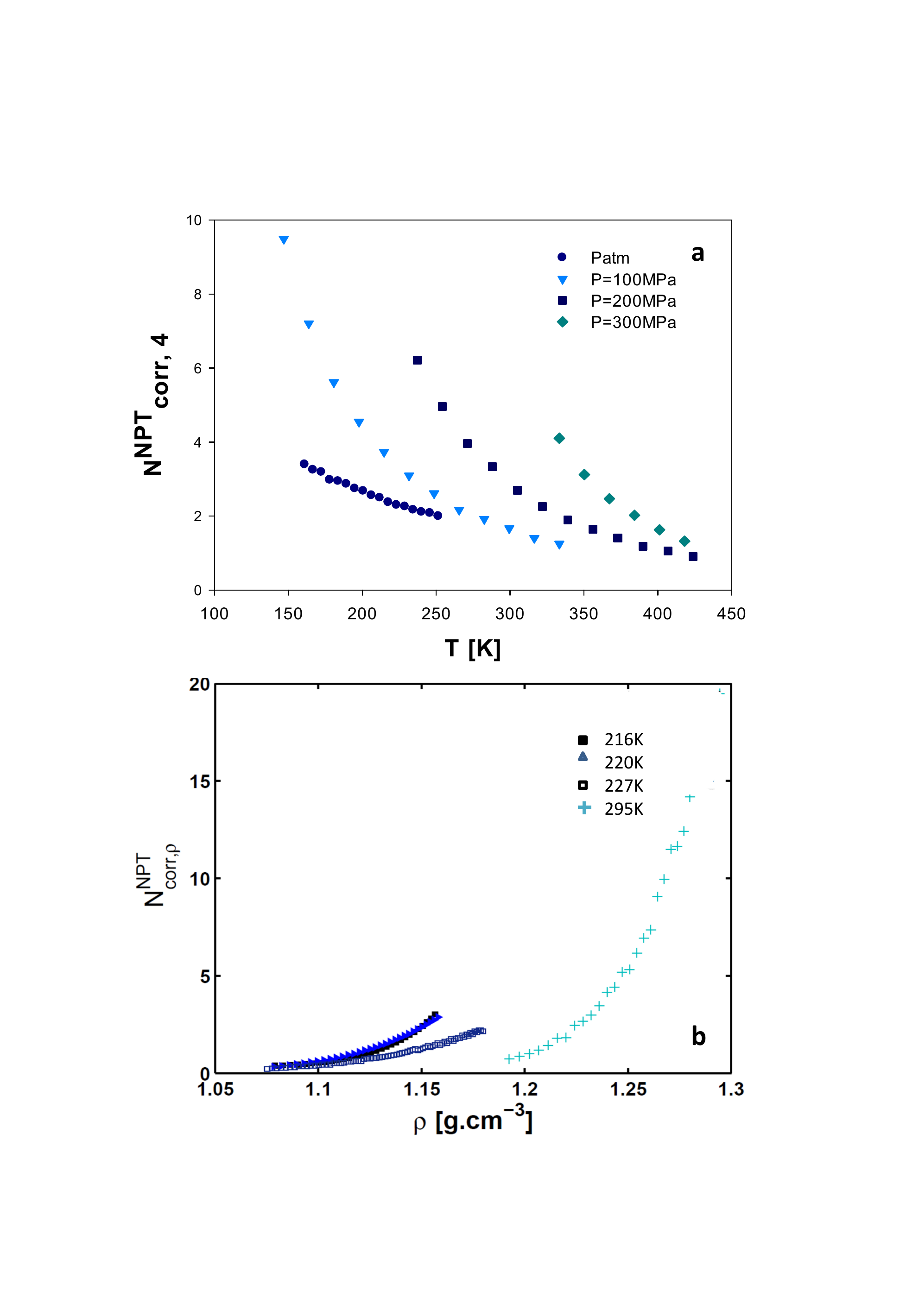}
\caption{ \label{fig2ab} \textbf{-a} Temperature dependence of the estimate of $N_{corr,4}^{NPT}$ from Eq. (\ref{eq:eqNcorrP_N})
along several isobars.
\textbf{-b} Density dependence of $N_{corr,\rho}^{NPT}$ along several isotherms : closed squares $T=216\,K$, closed triangles $T=220\,K$, open squares $T=227\,K$, crosses $T=295\,K$ (the data at $T=295\,K$
are from Ref. [\onlinecite{Paluch03}]). Both quantities
have been computed by a numerical differentiation of the Havriliak-Negami fits of the dielectric spectra. }
\end{figure}

We have first evaluated the four-point dynamical susceptibility along isobars through the simpler formula
in Eq. (\ref{eq:eqNcorrP}). To do so we have computed $\chi_T^{NPT}(t)$ from a numerical derivative of the
Fourier transform of the Havriliak-Negami fits of the experimental dielectric data, as explained in detail in Ref. [\onlinecite{Dalle07}].
We display the resulting estimate for $N_{corr,4}^{NPT}$ versus temperature $T$
on different isobars in Fig. \ref{fig2ab} (a). The results are similar to those obtained at atmospheric pressure
in previous papers \cite{Dalle07,Dalle08,Capaccioli08} and in Ref. [\onlinecite{Fragiadakis09}]: the number of dynamically correlated molecules is found to
increase when the temperature is decreased along any isobar, indicating growing spatial correlations
in the dynamics as one approaches the glass transition.

We have also considered the contribution coming from the density-triggered fluctuations along isotherms,
\textit{i.e.} $N_{corr,\rho}^{NPT}(\rho)$. To compute this term, we have numerically differentiated with
respect to density the Havrilliak-Negami fits (after converting the pressure data to density data) and included
the necessary thermodynamic factors. The results are shown in
Fig.\ref{fig2ab} (b).

 We observe that  $N_{corr,\rho}^{NPT}$ increases with increasing density along any isotherm. The other
contribution involving $\chi_T^{NVT}(t)=\left(\frac{\partial F(t)}{\partial \rho}\right)_T$ is virtually impossible to
get directly from the dielectric data with a good accuracy, due to the fact that
data are not collected along isochores. A different method will then be considered in the following section.
Note finally that the span of most curves in Figs. 2 (a) and (b) is quite restricted. In spite of the large number of state
points studied and the fine $T,P$ grid covered, the range of relaxation times that are directly accessible
experimentally on given isobars and isotherms remains limited, especially at low temperature and high pressure.

\section{$N_{corr}$ through the scaling description of the relaxation time}

We have already stressed that considering separately the fluctuation contributions due to density
and temperature provides a crisper estimate of the number of dynamically correlated molecules $N_{corr,4}^{NPT}$.
However, to compute the various terms required [see Eqs. (\ref{eq:eqNcorrrhoT}) and (\ref{eq:eqNcorrrhoT_N})], an
alternative procedure to the direct differentiation of the experimental spectra must be employed. We have
therefore relied on the, by now, well
established scaling description of the density and temperature dependences of the relaxation time.

\subsection{Density-temperature scaling law for relaxation time}

\begin{figure}%
\includegraphics*[width=\columnwidth]{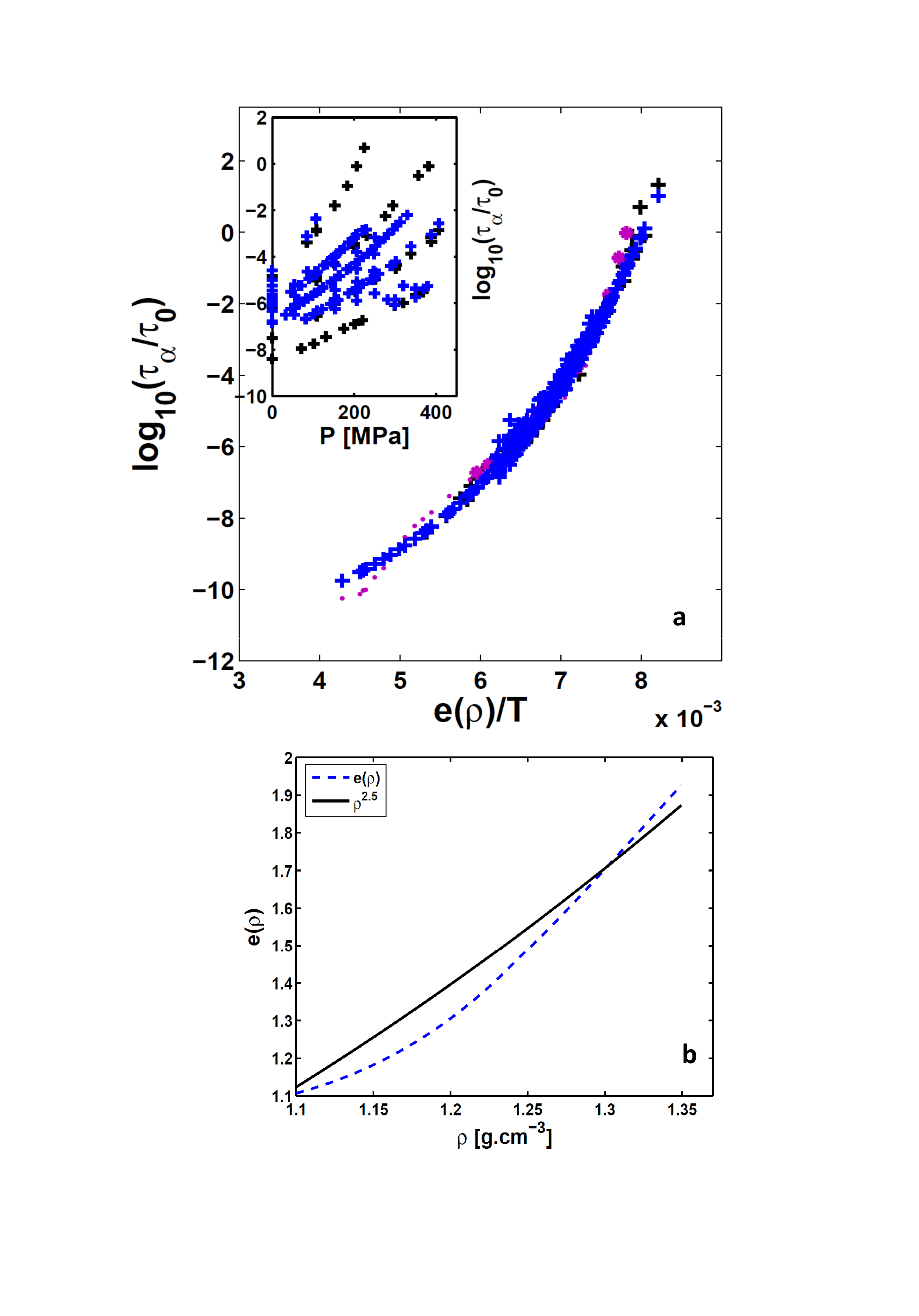}
\caption{ \label{fig3ab} \textbf{-a} Logarithm (base 10) of the (dielectric) relaxation time $\tau_\alpha$
(in units of $\tau_0=1 \rm{sec}$) of DBP measured at different pressures and temperatures
versus the scaling parameter $e(\rho)/T$. The latter is adjusted to obtain the best collapse of all data points.
The blue symbols denote our new set of data, the black ones our previous data from Ref. [\onlinecite{Niss07}],
and the violet ones are for literature data from Refs. [\onlinecite{Cook94,Paluch03}]. Inset: All our
data points plotted versus pressure.
\textbf{-b} Bare activation energy $e(\rho)$ obtained from the collapse of
relaxation time data (dotted line). For DPB the function cannot be simply described by a power-law
dependence $\rho^x$: the full line represents the best fit with a constant $x=2.5$. }
\end{figure}

The relaxation times $\tau_\alpha$ obtained from the Havriliak-Negami fits of the dielectric
spectra [$\tau_{\alpha}$ is equivalent to  $\tau$ in Eq. (\ref{eq:havriliak})] are plotted versus
pressure at different temperatures in the inset of Fig. \ref{fig3ab} (a). The main panel of this figure shows that the same
points, combined with literature data \cite{Cook93,Paluch03}, can be collapsed onto a master curve
when plotted versus a scaling parameter $e(\rho)/T$: the function $e(\rho)$ is obtained from
the best collapse of the whole data set and has the physical meaning (up to an overall constant factor) of a bare
or noncooperative activation energy characterizing the dynamics at high temperature \cite{Alba02}.
The fact that all the $\tau_\alpha$'s can be plotted on a master curve,
\begin{equation}
\label{eq_scaling_tau}
\log \left(\frac{\tau_\alpha(\rho,T)}{\tau_0}\right)=\mathcal F(\frac{e(\rho)}{T})\, ,
\end{equation}
with $\mathcal F$ a species-dependent scaling function, has been already proposed and verified for a great
variety of glass-forming liquids and polymers \cite{Alba02,Alba04,Roland04,Dreyfus04,Roland05}.
In many cases, the function $e(\rho)$ can be approximated by a power law, $\rho^x$, \cite{Alba04,Roland04,Dreyfus04}.
Often, however, the density range under study is too limited to allow a discrimination between
different functional forms for $e(\rho)$. Interestingly, in the case of DBP, data are available up to $1\,GPa$,
and it was found in Ref. [\onlinecite{Niss07}] that a simple power law does not provide a good collapse
of the data over the whole
density range and should be replaced by a more general function, \textit{e.g.} a
polynomial function. This is illustrated in Fig.\ref{fig3ab} (b). In such a case, the exponent $x$ can be generalized to
a density dependent parameter
\begin{equation}
\label{eq_scaling_x}
x(\rho)= \frac{\partial \log e(\rho)}{\partial \log (\rho)} \,,
\end{equation}
which varies between 1.5 and 4 for DBP over the range of density covered. As discussed
elsewhere\cite{Alba06} (and recently confirmed in the case of
``strongly correlating'' liquids \cite{Dyre}), there are no physical
reasons for why the density-temperature scaling law of the relaxation time should always
be expressible in terms of a simple power law. In this respect, the study presented in
this paper is more general than previous ones \cite{Fragiadakis09,Fragiadakis11}
that have only considered liquids and ranges of pressure for which the exponent $x$ can roughly be
taken as constant. We will see that this feature has some important consequences on
the variation of the number of dynamically correlated molecules.

\subsection{Density and temperature contributions to $N_{corr,4}^{NPT}$}

The density-temperature scaling of the relaxation time provides a continuous parametrization
of the latter over the whole domain of density and temperature and more specifically allows us
to compute derivatives with respect to the temperature at constant density, an otherwise horrendous
task experimentally. To further facilitate the computation of the various contributions to
$N_{corr,4}^{NPT}$, we have simplified the form of the relaxation function $F(t)$ by fitting the
result of the Havriliak-Negami parametrization to a stretched exponential,
$F(t)= \exp [-(t/\tau_{\alpha})^{\beta}]$, with $\beta$ a state dependent parameter.
This enables us to rewrite $N_{corr,T}^{NVT}$, $N_{corr,\rho}^{NPT}$, and $N_{corr,T}^{NPT}$ as
\begin{equation}
N_{corr,T}^{NVT}=\sqrt{\frac{k_B}{c_V}}\, \beta\, e^{-1}  \bigg [\frac{e(\rho)}{T} \mathcal F'(\frac{e(\rho)}{T}) \bigg ] \,
\label{eq:eqscaling_NcorrNVT}
\end{equation}
\begin{equation}
N_{corr,\rho}^{NPT}=\sqrt{\rho k_B T \kappa_T}\, \beta\, e^{-1}
x(\rho) \bigg [\frac{e(\rho)}{T} \mathcal F'(\frac{e(\rho)}{T}) \bigg ] \, ,
\label{eq:eqscaling_Ncorrrho}
\end{equation}
and
\begin{equation}
N_{corr,T}^{NPT}=\sqrt{\frac{k_B}{c_P}}\, \beta\, e^{-1} \bigg [1+x(\rho) T \alpha_P\bigg]
\bigg [\frac{e(\rho)}{T}\mathcal F'(\frac{e(\rho)}{T}) \bigg ],
\label{eq:eqscaling_NcorrNPT}
\end{equation}
where $\alpha_P$ is the thermal expansion coefficient, $\beta$ is the stretching parameter, $e^{-1}$ is the inverse of
Euler's mathematical constant, which should not
be confused with the activation energy $e(\rho)$, $x(\rho)$ is defined in eq. (\ref{eq_scaling_x}), and $\mathcal F'$ is the
first derivative of the scaling function in Eq. (\ref{eq_scaling_tau}). To derive the above equations, we have used the
fact that the maximum over time of the various susceptibilities calculated from a stretched exponential description occurs
for $t=\tau_{\alpha}$.

Finally, $N_{corr,4}^{NPT}$ is estimated from Eq. (\ref{eq:eqNcorrP_N}) as
\begin{equation}
N_{corr,4}^{NPT}\simeq (N_{corr,T}^{NVT})^2 (1+R),
\label{eq:eqNcorr_full}
\end{equation}
where $N_{corr,T}^{NVT}$ is given in Eq. (\ref{eq:eqscaling_NcorrNVT}) and $R$ is defined as \cite{Dalle07}
\begin{equation}
R=\rho T \kappa_T c_V \,x(\rho)^2 \,.
\label{eq:eqR}
\end{equation}
$R$ is therefore a measure of the relative importance of the
density-induced contribution to the number of dynamically correlated
molecules at constant pressure.

Before discussing the results obtained from Eq. (\ref{eq:eqNcorr_full}), it is worth assessing various
sources of uncertainty in the calculation. We have thus compared the values of $N_{corr,\rho}^{NPT}$
and of $N_{corr,T}^{NPT}$ computed from Eqs. (\ref{eq:eqscaling_Ncorrrho}) and
(\ref{eq:eqscaling_NcorrNPT}), respectively, to their counterparts obtained through a numerical derivative
of the Havriliak-Negami parametrization of the dielectric data (see previous section). We find that the difference
between the two sets are at most of the order of $10\, \%$ [which amounts to at most $20\,\%$ in the square values
entering in $N_{corr,4}^{NPT}$: see Eqs. (\ref{eq:eqNcorrP_N},\ref{eq:eqNcorrrhoT_N})], with no systematic
trend with density or temperature.

Additional errors come with the account of the thermodynamic input. First, there
can be a change of an overall factor for the absolute value of the $N_{corr}$'s according to whether one considers
the total fluctuations of energy, enthalpy or density (as rigorously enters in the lower bound for $N_{corr,4}^{NPT}$) or
only that part of the fluctuations which is associated with the structural relaxation and can be ascribed to the values
in excess to the glass ones (as physically motivated) \cite{Dalle07}. For instance, as described in section II-B,
$\Delta C_P$ is about $1/3$ of $C_P$ for DBP at the glass transition, which, when replacing $C_P$ by $\Delta C_P$ in
the estimate of the number of dynamically correlated molecules,  introduces an overall multiplying factor of $3$.
However, we are not interested in \textit{absolute} magnitudes
but in \textit{relative} variations with $T$, $\rho$, or $P$. Concerning the latter, the situation is much more favorable,
as one can estimate that the relative change in the various thermodynamic factors is never more than $10\,\%$
over the whole range of
state points under study. To summarize this discussion, we conservatively conclude that a variation of the estimate
of $N_{corr,4}^{NPT}$ that is beyond $20-30\, \%$ is physically significant. Smaller ones need to be interpreted with
caution.

\begin{figure}%
\includegraphics[width=\columnwidth]{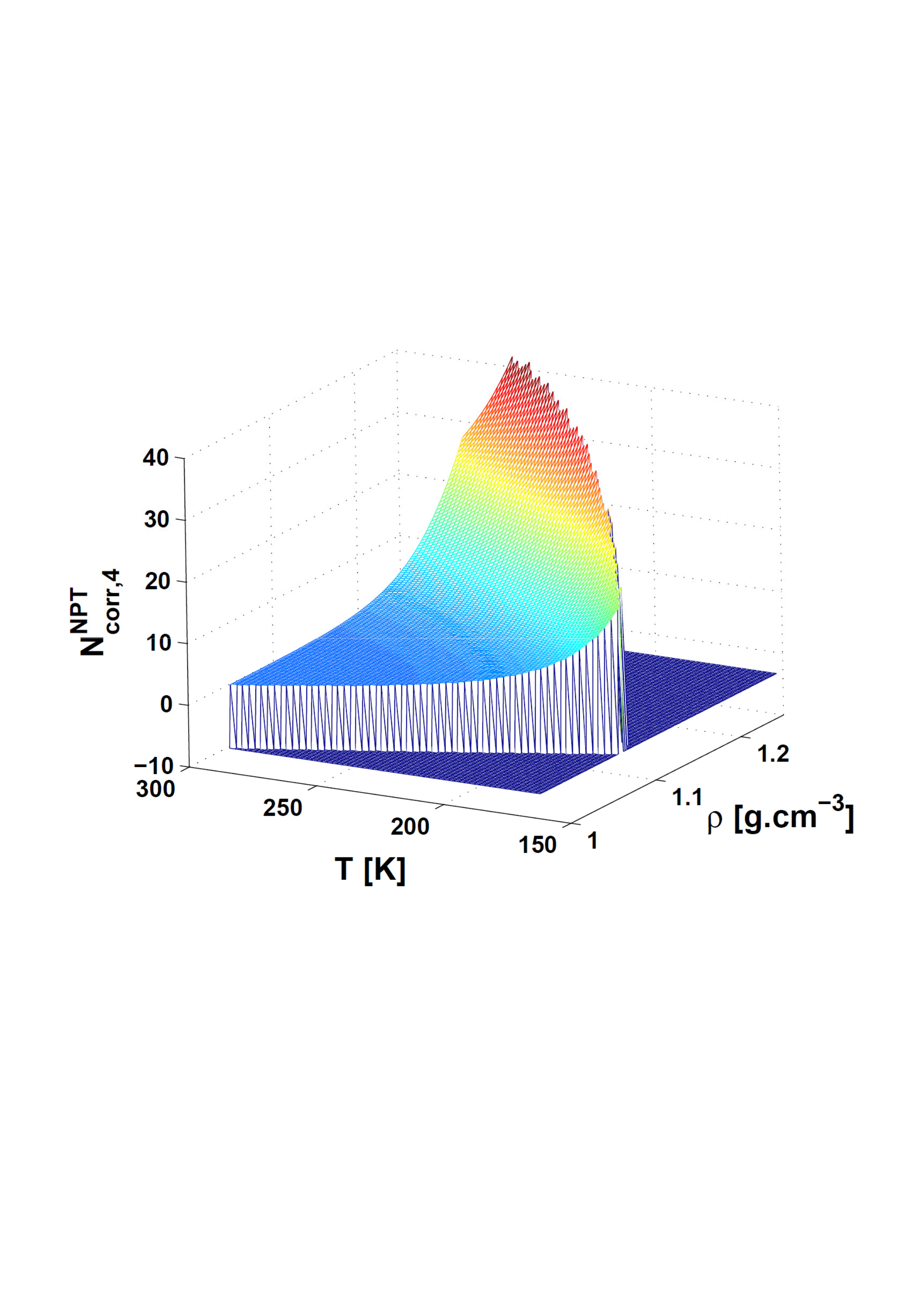}
\caption{ \label{fig4} 3D-plot for the number of dynamically correlated molecules $N_{corr,4}^{NPT}$ of DBP, estimated via Eqs.
(\ref{eq_scaling_tau}), (\ref{eq:eqNcorr_full}) and (\ref{eq:eqR}), for a wide range of $T$ and $\rho$.
$N_{corr,4}^{NPT}$ increases with increasing density and with decreasing temperature. The changes of color schematically
 indicate isochronic (constant relaxation time) conditions.}
\end{figure}

Fig. \ref{fig4} shows the variation with temperature and density of
the number of correlated molecules $N_{corr,4}^{NPT}$ in DBP, as obtained from
Eq. (\ref{eq:eqNcorr_full}) in conjunction with Eqs. (\ref{eq_scaling_tau},\ref{eq_scaling_x},
\ref{eq:eqscaling_NcorrNVT},\ref{eq:eqR}).
$N_{corr,4}^{NPT}$ is found to increase when the temperature decreases or
when density increases, as anticipated. The number of dynamically
correlated molecules obtained through this method is furthermore always slightly larger
than those calculated through the other, less efficient, bound in Eq. (\ref{eq:eqNcorrP_N})
or (\ref{eq:eqscaling_NcorrNPT}).

\section{$N_{corr,4}$ along the glass transition line and other isochrones}

\begin{figure}%
\includegraphics *[width=\columnwidth]{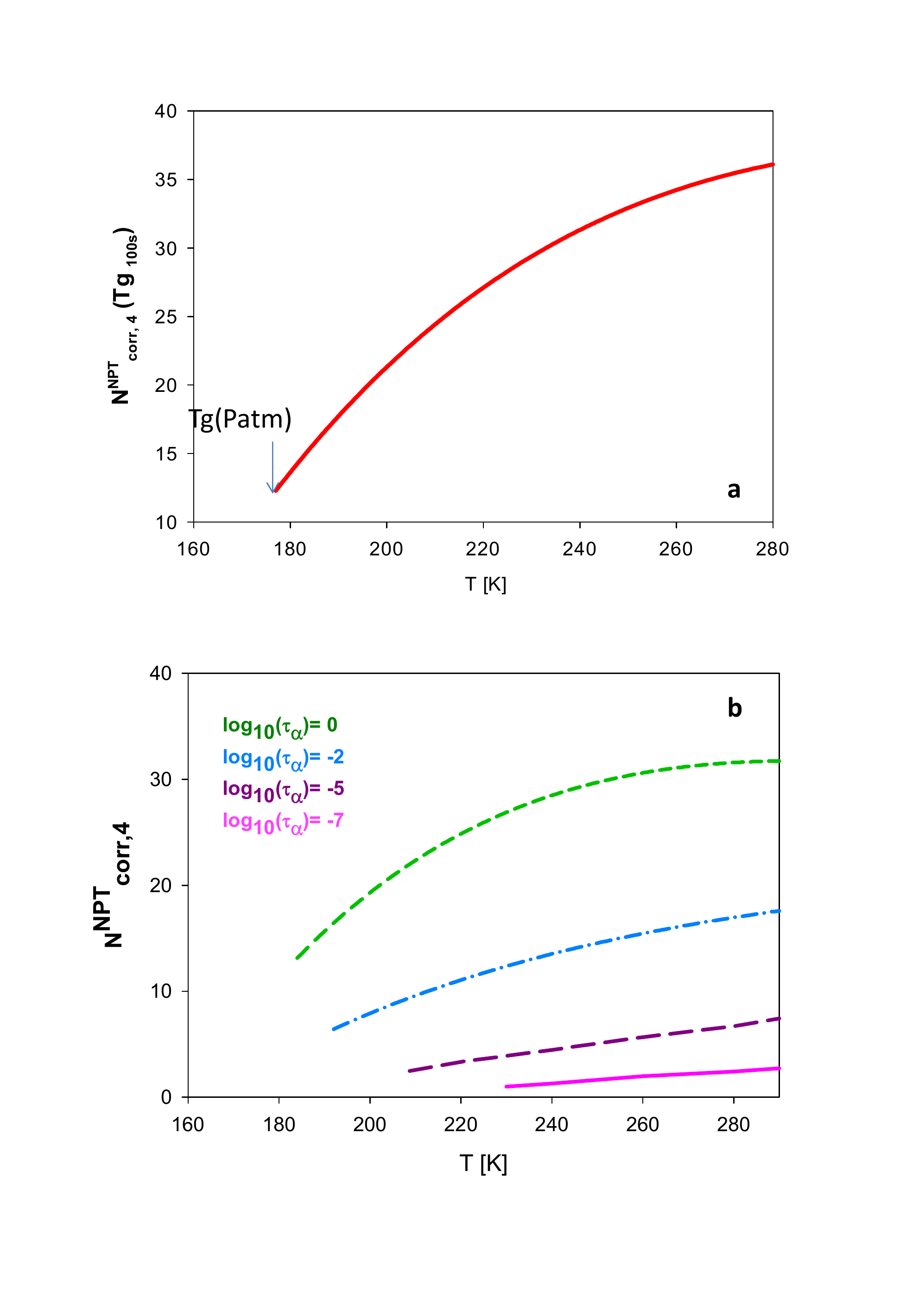}
\caption{ \label{fig5ab} \textbf{-a} Evolution of the number of dynamically correlated
molecules $N_{corr,4}^{NPT}$ along the glass transition line for DBP ($T_g$ is defined for a dielectric
relaxation of $100\, sec$). \textbf{-b} Same as (a) but for various isochronic lines
(the value of the $\log_{10}$ of the dielectric relaxation time in $sec$ is given in the caption)}
\end{figure}

From the 3-dimensional plot displayed in Fig. 4, one can follow the evolution of the number of dynamically
correlated molecules along any chosen path in the $(T,\rho)$ plane. An interesting choice of paths is
provided by the isochronic, \textit{i.e.} equal relaxation time, lines. The most commonly considered
among such lines is the glass transition line, where $T_g$ is defined by a given value of the relaxation
time, say $\tau_{\alpha}=100\,sec$. It was reported in Ref. [\onlinecite{Fragiadakis09}] for 4 glass-forming
liquids and polymers that the number of dynamically correlated molecules is uniquely determined
by the relaxation time. If this were true in general, $N_{corr,4}^{NPT}$ should be constant along
isochronic lines. As shown in Fig. \ref{fig5ab} (a), this is however not valid in the case of DBP. We find in
particular that $N_{corr,4}^{NPT}$ increases with pressure (and temperature) along the $T_g$ line,
at least up to the highest pressure available
%$1\,GPa$
(the evolution appears to saturate or reach a maximum for the highest pressures or temperatures considered).
The same systematic trend is observed for all isochrones, although the increase in $N_{corr,4}^{NPT}$
becomes smaller as the relaxation time decreases [see Fig. \ref{fig5ab} (b)]. We stress
that the pressure range covered in the present study on DBP is significantly larger than in
Ref. [\onlinecite{Fragiadakis09}]: here, we consider $P$ from the atmospheric value up to $1\,GPa$ (with a
change in density of  almost $25\,\%$) whereas $P$ varies from the atmospheric value to a few
tens to a few hundreds $MPa$ in [\onlinecite{Fragiadakis09}].

Two remarks are worth making in connection with the above result. First, the stretching parameter
$\beta$ of the relaxation functions was shown to be constant along any isochronic line, and in
particular along the $T_g$ line, in several glass-forming liquids \cite{Roland05,Niss07}. The same observation
applies here to liquid DBP. Secondly, the isochoric fragility $m_\rho$, defined as
$m_\rho=\left(\frac{\partial \log(\tau_\alpha/\tau_0)}{T\partial (1/T)}\right)_\rho$ is also found to be constant
along any isochronic line, as a result of the density-temperature scaling of the relaxation
time \cite{Alba02,Alba06}. The important consequence is that neither the stretching of the relaxation
nor the dynamic fragility are correlated with the increase in the number of dynamically correlated
molecules observed along the glass transition line and other isochrones.

\begin{figure}%
\includegraphics*[width=\columnwidth]{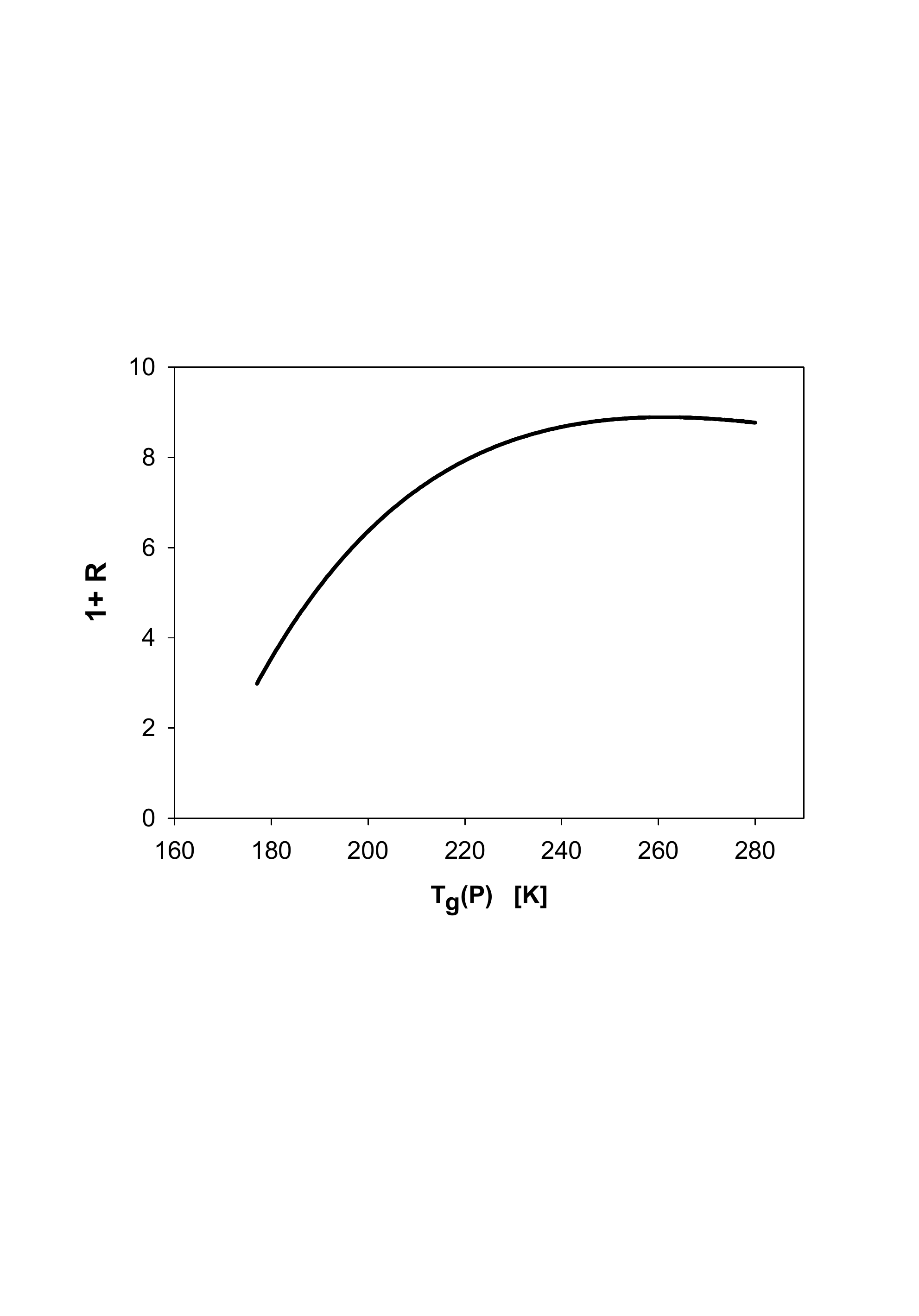}
\caption{ \label{fig6} Variation along the $T_g$ line of  $1+R$, where the parameter $R$
quantifies the relative importance of density-induced versus temperature-induced
contributions to $N_{corr,4}^{NPT}$. As for the number of dynamically correlated molecules, $1+R$ appears
to saturate or reach a maximum
 at the highest accessible pressures (and therefore temperatures)}
\end{figure}

As the stretching parameter $\beta$ and the isochoric fragility $m_{\rho}$ are essentially constant, whereas
$c_V$ varies by $10\,\%$ along the glass transition line, it follows from
Eqs. (\ref{eq:eqNcorr_full}) and (\ref{eq:eqscaling_NcorrNVT}) that the variation of
$N_{corr,4}^{NPT}$ with pressure should be mostly due to that of the parameter $R$. This is
indeed what is observed, as illustrated in Fig. \ref{fig6}: $1+R$ increases by a factor of almost 3 (and
$R$ by a factor of almost 4), which roughly corresponds to the increase in $N_{corr,4}^{NPT}$.

% regarder les options hauteur et angle 90° pour les figures
The physical
significance of this increase in $R$ is that the fluctuation effects triggered by density
become increasingly dominant over those associated with temperature as pressure and temperature
increase. We stress that this should not be confused with the relative importance of density effects (versus
temperature effects) in the slowing down of the relaxation. While $R$ concerns the \textit{fluctuations} around
the average dynamics, the latter deals with the \textit{average} dynamics (see the Introduction) and, in
the scaling description of the relaxation time, it is measured by the factor
$T \alpha_{P}(T,\rho)x(\rho)$ \cite{Alba06}. For DBP along the $T_g( P)$ line, $T \alpha_{P}$ is equal to
$0.12$
at atmospheric pressure
and decreases by about $25\,\%$ at high pressure,
% to $0.09$ at $1\,GPa$
whereas, as already quoted, $x$ varies from 2.5 to 4. As a result,
the factor $T \alpha_{P}x$ changes at the glass transition from $0.30$ to $0.36$ over the same pressure range.
At $T_g( P)$,  the influence of density on the average dynamics at constant pressure is thus significantly less
than that of temperature and increases by only $20\,\%$ under pressure,
a result very different from that found for $R$ in connection with the dynamical heterogeneities.

As the thermodynamic factor $\rho T \kappa_T c_V$ varies weakly with pressure along the glass
transition line, the increase in $R$ can be assigned to the change in the activation energy parameter
$x(\rho)$ [compare with Eq. (\ref{eq:eqR})]. The latter indeed increases from
$2.5$ to $4$ along the $T_g( P)$ line, which, once squared,
explains the growth of $R$ and of $N_{corr,4}^{NPT}$. The present study therefore
highlights the role of the deviation from simple power-law behavior of the bare activation energy $e(\rho)$
in the variation of the number of dynamically correlated molecules along the glass transition line. As a check,
we have repeated for another molecular glass-forming liquid, ortho-terphenyl, the very same
procedure as done here. This leads to a value of the number of dynamically correlated
molecules that increases along the glass transition line with increasing pressure up to $400\,MPa$,
but only by less than $20\,\%$. Accordingly,  it was found that the parameter $x(\rho)$ can be taken as
essentially constant over the same domain of pressure \cite{Dreyfus04,Alba02}.

Finally, we note that our finding (for DBP up to $1\,GPa$) of an increase of the number of dynamically correlated molecules,
and as a consequence of the associated correlation length, along the $T_g( P)$ line
is inconsistent with the theoretical predictions made on the basis of the Random First-Order Transition (RFOT)
theory \cite{Xia00,Lubchenko07}. In the RFOT theory, the dynamic length at $T_g$ is predicted to be
$\xi =5.8\,a$ \cite{Xia00,Lubchenko07}, where $a$ is the diameter of an elementary component of the
molecule or the polymer, referred to as a ``bead''. This prediction, which only depends on the relative
value of the relaxation time compared to a microscopic time, is claimed to be valid for all glass-formers
under any thermodynamic conditions. Some uncertainty do exist as to how beads should
be defined and operational procedures have been devised to compare different systems \cite{Xia00,Lubchenko07}.
However, when looking at the same liquid under different pressures, the bead should keep unchanged, which
sidesteps the fuzziness of its definition. The fact that $N_{corr,4}$ increases along the glass transition line
then directly contradicts the claim of the RFOT theory (except if one thinks that the relation between
length scale and number of molecules should change with pressure, an \textit{a priori} groundless assumption).
One may of course take the variation by a factor of slightly less than 3 which is observed observed for DBP
as ``within the noise'' of the theoretical predictions and the experimental procedures. Nevertheless, considering the care
with which the various sources of uncertainty have been estimated in our calculation, it rather seems  as
a fact to be seriously taken into account in further theoretical and experimental investigations.

\section{Conclusion}
In this work, we have characterized the spatial extent of the dynamical heterogeneities in a
glass-forming liquid through an experimental determination of the number of dynamically
correlated molecules obtained from dynamical susceptibilities. We have focused on the fragile liquid
dibutylphatalate (DBP), for which, from our own measurements and from existing ones, a very large domain
of temperature and pressure is experimentally covered for both dielectric spectroscopy and thermodynamics. We have
calculated different estimates of the number of dynamically correlated molecules $N_{corr,4}^{NPT}$, including
a crisp one making use of the separate contributions due to fluctuations associated with density and
with temperature. We have shown that the range of the spatial correlations in the dynamics of DBP
grow when approaching the glass transition along any thermodynamic path.
In addition, we have found that  $N_{corr,4}^{NPT}$ varies along the glass transition line (by a factor of about 3),
and more generally along any isochronic line, as pressure (or equivalently temperature) increases, a result which is at
odds with recent reports and theoretical predictions. The spatial extent of the correlations in the dynamics
is further found to be uncorrelated with both the (isochoric) fragility of the system and
the stretching of the relaxation function. This contradicts the naive view of the dynamics of glass-forming
liquids that ties cooperativity, fragility, stretching and growing heterogeneity altogether.
\\

\end{document}